\documentclass[aps,prb,twocolumn,superscriptaddress,floatfix]{revtex4-2}
\usepackage{amsmath}
\usepackage{amsfonts}
\usepackage{amssymb}
\usepackage{graphicx}
\usepackage{cancel}
\usepackage[switch,columnwise]{lineno}
\usepackage{titlesec}
\usepackage{xcolor}
\usepackage{hyperref}
\usepackage[none]{hyphenat}
\usepackage{braket}
\usepackage{mathrsfs}
\usepackage{footmisc}
\usepackage{float}
\usepackage{bbm}

\makeatletter
\renewcommand{\section}{\@startsection{section}{1}{0mm}
  {-\baselineskip}{0.5\baselineskip}{\bf\centering}}

\renewcommand{\subsection}{\@startsection{section}{1}{0mm}
  {-\baselineskip}{0.5\baselineskip}{\bf\centering}}
\makeatother

\begin{document}

\title{Discrete time quasi-crystal in Rydberg atomic chain}

\author{Xiaofan Luo}%
\affiliation{State Key Laboratory of Quantum Optics Technologies and Devices, Institute of Opto-electronics, Shanxi University, Taiyuan, Shanxi 030006, China}%

\author{Yaoting Zhou}%
\affiliation{State Key Laboratory of Quantum Optics Technologies and Devices, Institute of Opto-electronics, Shanxi University, Taiyuan, Shanxi 030006, China}%

\author{Zhongxiao Xu}%
\email{xuzhongxiao@sxu.edu.cn}
\affiliation{State Key Laboratory of Quantum Optics Technologies and Devices, Institute of Opto-electronics, Shanxi University, Taiyuan, Shanxi 030006, China}%
\affiliation{Collaborative Innovation Center of Extreme Optics, Shanxi University, Taiyuan, Shanxi 030006, China}%

\author{Weilun Jiang}%
\email{wljiang@sxu.edu.cn}
\affiliation{State Key Laboratory of Quantum Optics Technologies and Devices, Institute of Opto-electronics, Shanxi University, Taiyuan, Shanxi 030006, China}%
\affiliation{Collaborative Innovation Center of Extreme Optics, Shanxi University, Taiyuan, Shanxi 030006, China}%

\begin{abstract}
Discrete time quasi-crystals are non-equilibrium quantum phenomena with quasi-periodic order in the time dimension, and are an extension of the discrete time-crystal phase. As a natural platform to explore the non-equilibrium phase of matter, Rydberg atomic arrays have demonstrated the quantum simulation of the discrete-time crystal phase, associated with quantum many-body scar state.
However, the existence of discrete time quasi-crystal on the Rydberg cold atom experiment platform has yet to be conceived.
Here, we propose a method to generate the discrete time quasi-crystal behavior by coupling two discrete time-crystals, where associated two external driving frequencies have the maximum incommensurability.  
While analyzing its robustness and computing the phase diagram of corresponding observables, we significantly calculate the entanglement entropy between two parts of the system. Remarkably, we find the emergence of the aperiodic response is indeed caused by interaction between systems via Rydberg blockade effect. Our method thus offers the possibilities to explore the novel phases in quantum simulator.

\end{abstract}

\maketitle

\section{Introduction}

The quest for exotic quantum phases and novel dynamical regimes has driven remarkable advances in quantum simulation platforms. Concurrently, the concept of time crystals (TCs) — a non-equilibrium phase of matter characterized by spontaneously broken time-translation symmetry, proposed by Wilczek 
\cite{wilczekQuantumTimeCrystals2012}, has expanded the frontier of many-body physics. Unlike conventional crystals that break spatial symmetry, TCs manifest the periodicity in time, thus related to the many-body dynamics and non-equilibrium physics \cite{zaletelColloquiumQuantumClassical2023,carraro-haddadSolidstateContinuousTime2024,estarellasSimulatingComplexQuantum2020,freyRealizationDiscreteTime2022,liuPhotonicMetamaterialAnalogue2023,lyubarovAmplifiedEmissionLasing2022,raskatlaContinuousSpaceTimeCrystal2024,tragerRealSpaceObservationMagnon2021,wangMetasurfacebasedRealizationPhotonic2023,ieminiBoundaryTimeCrystals2018, leeAntiferromagneticPhaseTransition2011,pizziHigherorderFractionalDiscrete2021,russoQuantumDissipativeContinuous2025, solankiChaosTimeDissipative2024,wangBoundaryTimeCrystals2025,lourencoGenuineMultipartiteCorrelations2022,fanDiscreteTimeCrystal2020}. Over the past decades, experimental breakthroughs have transformed this theoretical curiosity into reality, with remarkable progress achieved across diverse platforms, such as the trapped ions \cite{zhangObservationDiscreteTime2017,kyprianidisObservationPrethermalDiscrete2021}, atom-cavity system\cite{kesslerObservationDissipativeTime2021,kongkhambutObservationContinuousTime2022a}, Nitrogen-vacancy (NV) centers \cite{choiObservationDiscreteTimecrystalline2017,randallManybodylocalizedDiscreteTime2021,heExperimentalRealizationDiscrete2025}, hot Rydberg vapor \cite{wuDissipativeTimeCrystal2024,liuHigherorderFractionalDiscrete2024,jiaoManybodyNonequilibriumDynamics2024,jiaoObservationTimeCrystal2024,jiaoQuantumLotkaVolterraDynamics2024}, and Rydberg atoms chain \cite{Controlling,DTC}, etc. 

\begin{figure*}
\centering
\includegraphics[width=1\textwidth]{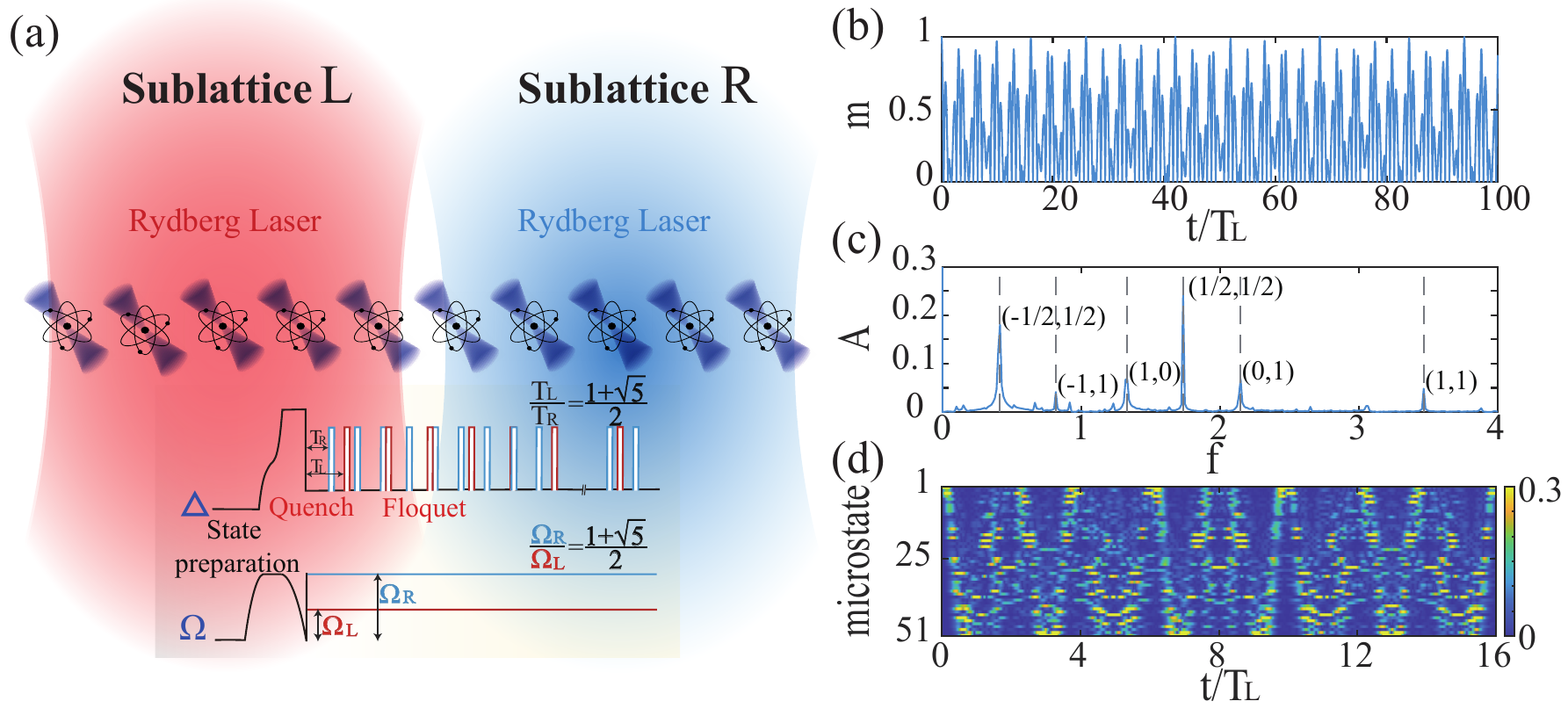}
\caption{\label{Fig:setup} Experimental proposal and numerical results. (a) Experimental setup. An one-dimensional atomic chain is divided into left and right subsystems. At the state preparation process, we prepare the atomic chain in the $Z_2$ state. And then, we conduct the quench process by rapidly adjusting the detuning $\Delta$ to 0, and transferring the Rabi frequency ratio of the left and right subsystems to $\Omega_{R}/\Omega_{L}=(\sqrt{5}+1)/2$. Meanwhile, we apply Floquet periodic pulse modulation separately to the left and right sides, and the modulation period satisfies $T_{L}/T_{R}=(\sqrt{5}+1)/2$. The sequence of events are shown in the lower half of the schematic diagram. (b) $m$ as a function of time after quenching, with the time as the unit of $T_L$. Here we set $N=32$, $\Omega_{L}=1$, $T_{L}=4.74$, $f_{L}=2\pi/T_{L}$, $\theta = \pi$. 
(c) The frequency spectrum of $m$.  $A$ represents for amplitude of frequency. 
The typical peaks are labeled with $(k_1, k_2)$. (d) The evolution trajectory of the quantum state in the  Hilbert space after quenching. We set $N = 9$, $N_L = 5$, and $N_{R} = 4$, thus the state arrangement on the vertical axis is same with that in Ref. \cite{Controlling} (51 microstates). The color depth of the heatmap indicates the overlap $\left <\psi_{n}| \Psi(t) \right>$ between the system and the basis states on the vertical axis. And the horizontal axis starts with the $Z_2$ state at the top and ends with the $Z_{2}^{\prime}$ state at the bottom.}
\end{figure*}

Among various system, the Rydberg atoms, characterized by their highly excited electronic states, exhibit exceptional properties such as strong, tunable dipole-dipole interactions and microsecond-scale coherence times \cite{giudiciFastEntanglingGates2024,jakschFastQuantumGates2000,lukinDipoleBlockadeQuantum2001a,zhangSubmicrosecondEntanglingGate2020}. These features enable the realization of programmable quantum simulators capable of emulating spin models \cite{blockScalableSpinSqueezing2024,bornetScalableSpinSqueezing2023,chenContinuousSymmetryBreaking2023,ebadiQuantumPhasesMatter2021,emperaugerBenchmarkingDirectIndirect2025,emperaugerTomonagaLuttingerLiquidBehavior2025,geierFloquetHamiltonianEngineering2021,labuhnTunableTwodimensionalArrays2016,maskaraProgrammableSimulationsMolecules2025,qiaoRealizationDopedQuantum2025,schollQuantumSimulation2D2021,slagleQuantumSpinLiquids2022,warmanCategoricalSymmetriesSpin2025}, topological states \cite{deleseleucObservationSymmetryprotectedTopological2019,giudiceTrimerStates32022,giudiciDynamicalPreparationQuantum2022,semeghiniProbingTopologicalSpin,weberExperimentallyAccessibleScheme2022,xieChiralSwitchingManybody2024}, and out-of-equilibrium phenomena \cite{Probing,celiEmergingTwoDimensionalGauge2020,bluvsteinQuantumProcessorBased2022,cesaUniversalQuantumComputation2023,choiPreparingRandomStates2023,ebadiQuantumOptimizationMaximum,huangObservationManyBodyQuantum2021,keeslingQuantumKibbleZurek,manovitzQuantumCoarseningCollective2025,zhangQuantumSlushState2024}, thus it emerges as a paradigmatic system for exploring many-body physics. Most notably, the strong interaction between Rydberg excitations brings about the famous Rydberg blockade effect, making it distinguished from the other platforms. Such a mechanism acts directly on the Hilbert space by restricting the adjacent excitations, and further leads to slow thermalization of the system and offers the possibility for realizing the TC behavior. Specifically, in the eigenvalue spectrum, few special states have relatively long thermalization time, known as the quantum many-body scar states (QMBS) states \cite{Probing,hoPeriodicOrbitsEntanglement2019,turnerQuantumScarredEigenstates2018,turnerWeakErgodicityBreaking2018,serbynQuantumManybodyScars2021,suObservationManybodyScarring2023a,halimehRobustQuantumManybody2023,windtSqueezingQuantumManyBody2022}, which largely overlap with the antiferromagnetic product states. To control its dynamics, recent experimental breakthroughs have demonstrated the potential of the Floquet-engineered approach.  For the Rydberg system, the precise control over atomic positions and laser-driven excitations allows the implementation of tailored Floquet protocols \cite{giudiciUnravelingPXPManyBody2024,michailidisSlowQuantumThermalization2020,restrepoDrivenOpenQuantum2016,rovnyObservationDiscreteTimeCrystalSignatures2018,zhangDigitalQuantumSimulation2022,zhaoFloquettailoredRydbergInteractions2023,bukovUniversalHighFrequencyBehavior2015,hudomalDrivingQuantumManybody2022,mizutaExactFloquetQuantum2020,mukherjeeCollapseRevivalQuantum2020}. In particular, Lukin et al. recently reported the realization of Floquet engineered discrete time crystal (DTC), where the rigid subharmonic temporal response appears under periodic driving \cite{Controlling,DTC}. 

Along this line, we are curious if one can further realize discrete time quasi-crystal (DTQC), as an extension of the DTC. Refer to the solid-state physics, the emergence of DTQC should fulfill two requirements: (1) Its symmetry should be lower than the external drive, which means the response further breaks the time-translation symmetry with respect to the drive. This feature is identical to the DTC phase; (2) An oscillatory response is formed, but without an evident period \cite{heExperimentalRealizationDiscrete2025,auttiObservationTimeQuasicrystal2018,elseLongLivedInteractingPhases2020a,pizziPeriodDiscreteTime2019,zhaoFloquetTimeSpirals2019}. Although the DTQC has been observed in NV centers very recently 
 \cite{heExperimentalRealizationDiscrete2025}, the protocol for realizing the DTQC in Rydberg atoms chain has yet to be conceived.

In this paper, we propose a scheme of the DTQC in one-dimensional Rydberg atomic array driven by external time-modulated fields. Given the properties of the QMBS, we investigate how emergent quasiperiodic temporal order can be stabilized. We define the observables, such as the antiferromagnetic order parameter, to describe the DTQC phase, and find the associated time evolutions exhibit oscillations but aperiodic. With Fourier spectra, we validate the existence of the DTQC and obtain its phase diagram. Significantly, the results of the entanglement entropy reveal the DTQC phase can only exist under the proper driving frequency. Out of this parameter region, the higher frequency could leads to the decouple between two subsystems, while the lower frequency will result in the chaotic signal of the response. 

\section{Model}

We begin by introducing the theoretical description of Rydberg atoms chain. The strong interaction between atoms ensures that there are no adjacent Rydberg excitations, named Rydberg blockade effect. Such effect of Rydberg chains is well captured by the PXP model \cite{lesanovskyInteractingFibonacciAnyons2012}. In the one-dimension chain, QMBS emerges as typical characters, which expresses the ergodicity broken behavior in the time-evolution dynamics. This further results in the fact that the few initial states, which have relative large overlap with QMBS, could exhibit the periodic recovery phenomenon after quench, i.e. form a closed evolution track in the Hilbert space \cite{Probing}. And its frequency depends on the Rabi frequency of the quenched Hamiltonian. However, the system will finally reach the thermal equilibrium after a long-time evolution. In order to prolong the existence time of the recovery, the Floquet modulations are employed in the experiment. Specially, the oscillation is stabilized with its frequency equal to the half of the external driving frequency, thus the system shows the behavior of the DTC \cite{Controlling}. In fact, to form DTC phase, the driving frequency is only feasible within a certain range according to the inherent oscillation frequency of the PXP model, which is determined by the energy differences between the scar states. 

On this basis, to construct a DTQC phase referred to the properties of DTC and QMBS, a possible approach is to divide the atomic chain into two parts, denoted by $L$ and $R$. The purpose of bipartite lattice structure design is to directly connect the two incompatible DTCs via blockade effect. Concretely, we set the different Rabi resonant frequency for the $L$-subsystem and the $R$-subsystem as $\Omega_{L}$ and $\Omega_{R}$, and different pulse sequence period $T_{L}$ and $T_{R}$, respectively. The static Hamiltonian of the model writes as follows:
\begin{equation}\label{PXPHamiltonian} 
\hat{H}_{\text{PXP}} = \sum_{i\in{L}} \frac{\Omega_{L}}{2}\hat{P}_{i-1}\hat{X}_{i}\hat{P}_{i+1} + \sum_{i\in{R}} \frac{\Omega_{R}}{2}\hat{P}_{i-1}\hat{X}_{i}\hat{P}_{i+1},  
\end{equation}
where $\hat{P}_{i} = \ket{g_{i}}\bra{g_{i}}$ denotes the projection operator of the ground state, $\hat{X}_{i} =\ket{r_{i}}\bra{g_{i}} + \ket{g_{i}}\bra{r_{i}}$ represents the transit of the atom between the ground state and the Rydberg state on site $i$. To mimic the experiment, we use the open boundary conditions (OBC) for our computations.

Moreover, to realize the quasi-periodic behavior, we set the ratio between $\Omega_{R}$ and $\Omega_{L}$ to be the maximum incommensurate ratio $r= (\sqrt{5} + 1)/2$. Considering that the inherent oscillation frequency is in linear proportion to the Rabi frequency \cite{turnerQuantumScarredEigenstates2018,turnerWeakErgodicityBreaking2018}, we hence set the ratio of two Floquet pulse modulation satisfying $T_{L} / T_{R} = \Omega_{R}/ \Omega_{L} = r$ to stabilized the coupled system. As for the modulation pattern, we refer to the Ref. \cite{DTC} and use $\delta$-functions as an example. 
Therefore, the Floquet modulation Hamiltonian is,
\begin{equation}\label{FloquetHamiltonian}
\hat{H}_{\text{F}} = \theta_{L}\hat{n}_{L} \sum_{k_{L}\in{\mathbb{Z}}}\delta(t-k_{L}T_{L}) + \theta_{R}\hat{n}_{R} \sum_{k_{R}\in{\mathbb{Z}}}\delta(t-k_{R}T_{R}),
\end{equation}
where $\hat{n}_{L/R} = \sum_{i\in{L/R}} \ket{r_{i}}\bra{r_{i}}$, represents the number of particles on the Rydberg state of the $L$-subsystem or $R$-subsystem, and $\theta$ is the modulation strength. For simplicity, we set $\theta_{L} = \theta_{R} = \theta$ for following computation if not specified. Thus, the total Hamiltonian of the system is $\hat{H}(t)=\hat{H}_{\text{PXP}} +\hat{H}_{\text{F}}(t)$, as shown in Fig. \ref{Fig:setup}(a). The only tunable parameters are Rabi frequency $\Omega_L$, modulation period $T_L$ (or denoted as driving frequency $f_L = 2\pi/T_L$), and modulation strength $\theta$. Here we set $\Omega_L = 1$ as the unit. 

In the following simulations, we use the diagonalization method to solve the system of size $N < 14$, and the simulation time step is chosen as $dt = 0.001$. For larger sizes, we use the time-dependent variational principle based on the matrix product state, with the truncation error lower than $\epsilon = 5 \times 10^{-9}$, and the maximum bond dimension is set as 500 \cite{hauschildEfficientNumericalSimulations2018,haegemanUnifyingTimeEvolution2016,haegemanTimeDependentVariationalPrinciple2011}.

\section{Discrete time quasi-crystal}

As pointed out by Ref. \cite{DTC}, under the system Hamiltonian $H(t)$, the $Z_2$ state ($\ket{\bullet\circ\bullet\circ...\bullet\circ}$,  where $\bullet$ means excited atom and $\circ$ means atom on the ground state) evolutes toward its analog $Z_2^\prime$ state ($\ket{\circ\bullet\circ\bullet...\circ\bullet}$), and vice versa. The period for the reversion to the initial state is twice of the modulation period. As a result, the 2-order DTC is demonstrated by the defined observables, which is relevant to the properties of the initial state. In this work, we put forward such an observable for describing the time crystal behavior, whose definition follows the antiferromagnetic order parameter $m$,
\begin{equation}\label{observable}
    m = \frac{|\sum_{i} (-1)^i \hat n_i|}{N}.
\end{equation}
Here $\hat n_i = \ket{r_{i}}\bra{r_{i}} $ is Rydberg population density on site $i$. Obviously, the initial $Z_2$ state has maximum value of $m = 1$, it is thus denoted as the antiferromagnetic state. And the observable provides a good description of the oscillation behavior between the two antiferromagnetic states $Z_2$ and $Z_2^\prime$. 
 
Here, we perform the simulation by assigning $Z_2$ state  as the initial state,
we find that at appropriate parameter range, the observable $m$ shows a stable recovery phenomenon in Fig.\ref{Fig:setup}(b), in contrast to the rapid thermalization without the Floquet drive as depicted in the Appendix A. However, its period is not obvious and hence not easily to extract. We thus use the Fourier transformation to analyze its spectrum. As mentioned, the peaks in the Fourier spectrum of 2-order DTQC phase should be subharmonic and incommensurate. In practice, the different frequencies may be mixed with each other. Therefore, the expected frequency response $f$ is formalized as,
\begin{equation}\label{frequencyfunction}
f(f_{L},f_{R})=\frac{k_{1}}{2} \times f_{L}+\frac{k_{2}}{2} \times f_{R},\qquad k_{1},k_{2}\in{\mathbb{Z}}.
\end{equation} 
Note we employ $(k_1, k_2)$ to label the different peaks. 
For our simulation, we numerically observe several sharp peaks in Fig.\ref{Fig:setup}(c) after a long-time evolution, corresponding to the complicated aperiodic behavior. Moreover, we find the frequency responses are mainly concentrated at the position (1/2, 1/2), and (-1/2, 1/2), which are the summation and subtraction of the halve of two incommensurable driving frequencies. Such behavior accords with the definition of the DTQC in phenomenon. 

To clearly show the evolution track among various states, in Fig. \ref{Fig:setup}(d), we draw the time evolution diagrams of the unrestricted product states $\ket{\psi_{n}}$, which obey the Rydberg blockade rule. Similar to DTC phase in Ref. \cite{DTC}, the bipartite modulations in our setup also deepen the scar trajectory and delay the thermalization of the system. By contrast, when the system is initialized on $Z_2$ state without modulation, there is completely no periodic recovery phenomenon of the whole system $H_{\text{PXP}}$, and the system will quickly thermalize. This can be seen from the comparison of dashed and solid black lines in Fig. \ref{Fig:setup}(b).  The reason is that the bipartite construction and incommensurate coupling of our model, which destroys the scar states. Nonetheless, under the maximum incommensurate Floquet double modulation, the system still exhibit stable oscillation between the $Z_2$ and $Z_2^\prime$ state, even though the period of the state evolution trajectory becomes less obvious.

\begin{figure}
\centering
\includegraphics[width=0.5\textwidth]{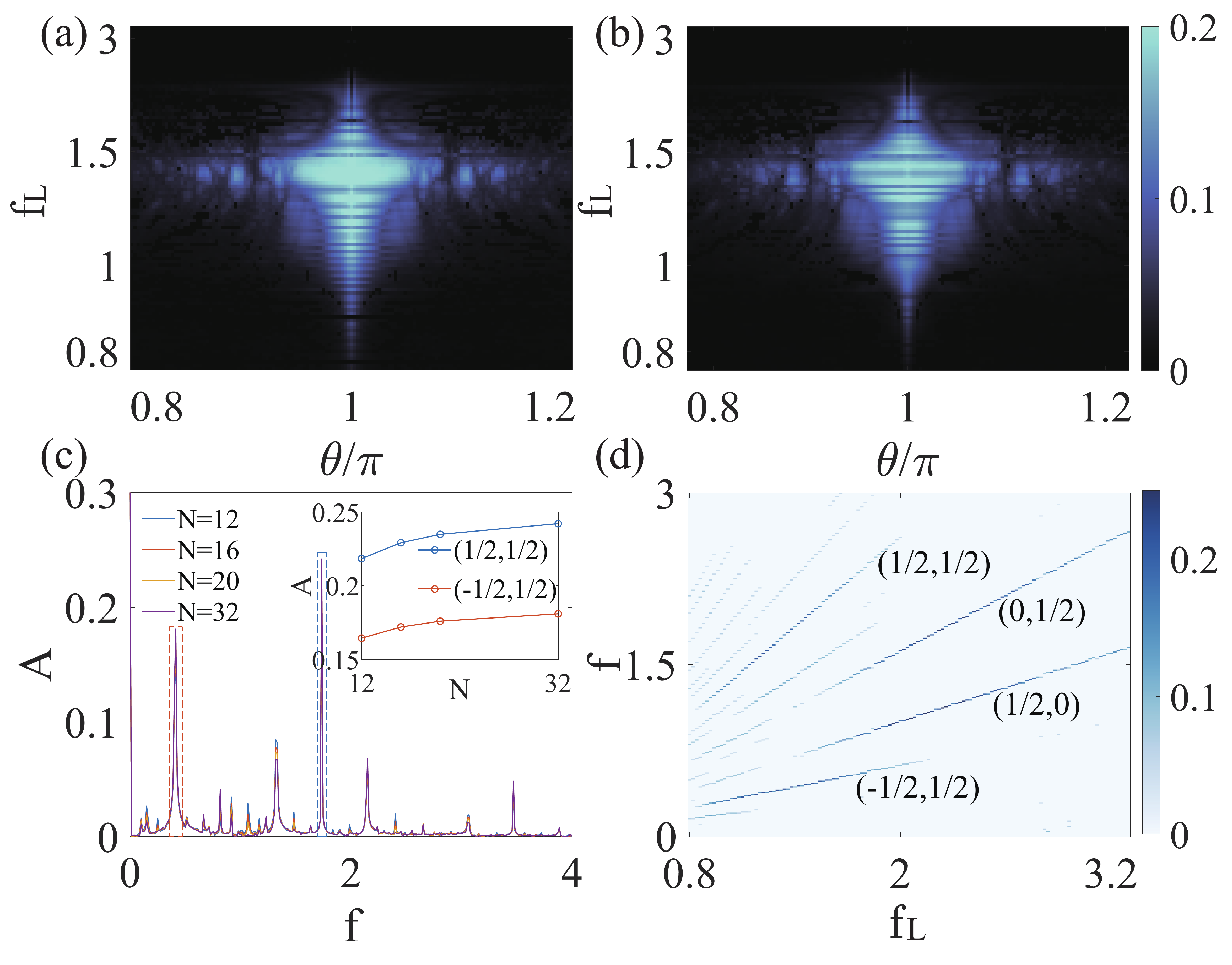}
\caption{\label{Fig2} The phase diagram and its parameter dependence. (a, b) The phase diagrams of DTQC of modulation intensity $\theta$ and the modulation frequency $f_{L}$. The brightness indicates its amplitude $A$ of the peaks with respected to the frequency components for (-1/2, 1/2) in (a) and (1/2, 1/2) in (b). We set $N = 10$ and the evolution time $t = 1000$ for extracting the Fourier spectrum of observable $m$. We suppose the system is in the DTQC phase, on the condition of the oscillatory lifetime of frequency $\tau > 30 T_{L}$.
(c) The size-dependence of the of typical peaks (-1/2, 1/2) and (1/2, 1/2). (d) The variation of spectral components with the driving frequency $ f_{L}$. The brightness indicates its amplitude $A$ of the peaks. }
\end{figure}

\section{Phase diagram}

Since DTQC exists as a phase in the parameter space, we thus design an appropriate observable to describe the phase region. Notice that the long-time evolution of the antiferromagnetic observable is asymptotic to expressed as 
\begin{equation}\label{FloquetHamiltonian}
m \approx \sum_{f}A_{f} e^{-\frac{t}{\tau_{f}}}\cos(ft),
\end{equation}
where $\tau_f$ represents the lifetime of the component with frequency $f$, and $A_{f}$ denotes the amplitude of oscillation. Fig. \ref{Fig2}(a) and (b) show the phase diagram of the DTQC phase with respect to the frequency component (1/2,1/2) or (-1/2,1/2). The phase regions of both components are almost the same. In addition, we observe that the phase region becomes broad near the condition $\theta = \pi$ and $f \sim 1.4$. On the one hand, $\theta = \pi$ is just the parameter choice for implementing the DTC on each bipartite \footnote{ Considering only the left part, the Hamiltonian defined on the left part $\hat{H}_{\text{PXP,L}}$ anticommutes with the Floquent operator on the left part 
$\hat{\mathcal{C_L}}=e^{-i\pi\hat{n_{L}}}$. The operation for doubling acting the time evolution operator is exactly equal to the unit operator $\hat{\mathcal{C}}e^{-iT_{L}\hat{H}_{\text{PXP,L}}}\hat{\mathcal{C}}e^{-iT_{L}\hat{H}_{\text{PXP,L}}}=e^{iT_{L}\hat{H}_{\text{PXP,L}}}e^{-iT_{L}\hat{H}_{\text{PXP,L}}}=\mathbbm{1}$.}. That is, after two modulation periods, the subsystem will exactly returns to its initial state and implies perfect subharmonic revivals, which is valid for both bipartite \cite{DTC}. We expect such results could be generalized to the DTQC. Therefore, the greater the deviation of the modulation intensity from $\pi$, the worse the revives of the system will be. 
On the other hand, $f \sim 1.4$ is closed to the double of the inherent oscillation frequency of the $L$-subsystem, where the inherent oscillation associated with $\Omega_L$ is approximately 0.7 as in Fig. \ref{Fig:setup}(b). The effect of Floquet modulation performs better around such frequency than others.

To further demonstrate the robustness of DTQC phase, we explore its size and frequency dependence. In Fig. \ref{Fig2}(c), we plot the size dependence of the spectrum for the observable $m$. As the system grows, the amplitudes of the typical peak (-1/2, 1/2) and (1/2, 1/2) approach a constant. This indicates that the finite size effect is not severe in the observable for such subharmonic response, which is consistent with the findings in Ref. \cite{Controlling}.  Fig. \ref{Fig2}(d) shows the frequency response of the system at the modulation strength $\theta = \pi$, by varying the modulation frequency $f_{L}$. As expected, all the responses are linearly proportional to the driving frequency. At small frequency limit, plenty of Fourier peaks appear, including the component of the original inherent oscillation, i.e. the peak labeled (1,0). At this time, the system is fully chaotic, instead being regarded as a crystal in time domain. Conversely at large frequency limit, only (-1/2,0) and (0,1/2) survives. At this time, the whole system turns into decoupled bipartite, where the response is the simple additivity of two independent DTCs. When we choose $f_L$ in an intermediate range, the sharp (-1/2, 1/2) and (1/2, 1/2) peaks will dominant, in agreement with the criteria of Eq. \eqref{frequencyfunction}. In such region in the phase diagram, we suppose the system behaves like a good DTQC. In Appendix B, we also investigated the dependence of the system's initial state.


\begin{figure}[t]
\centering
\includegraphics[width=0.45\textwidth]{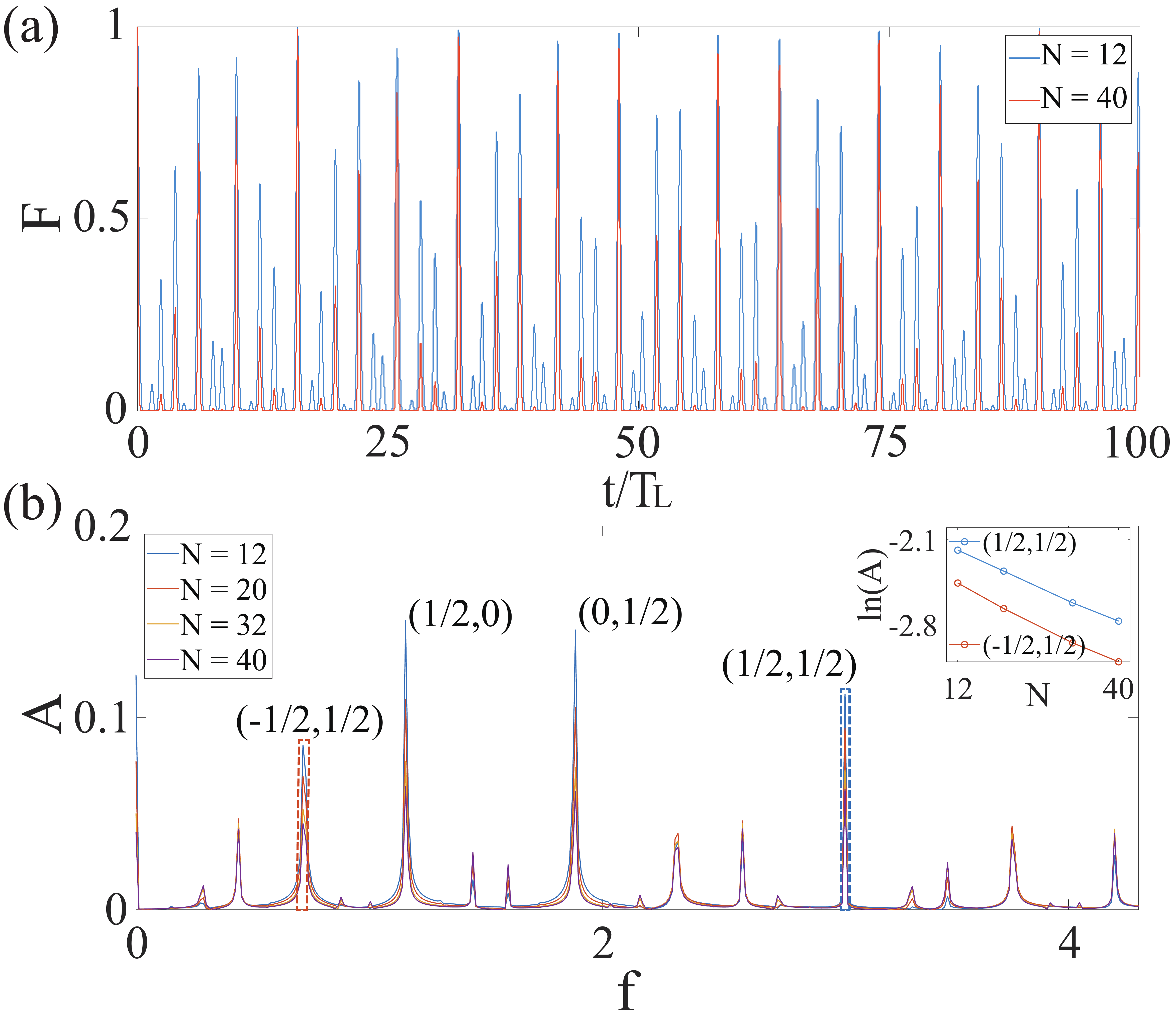}
\caption{\label{Fig3} The results of the fidelity $\langle Z_2| \psi (t)\rangle$. (a) Fidelity as a function of time. $T_{L} = 2.32 $. (b) Its associated Fourier spectrum. We performed Fourier transformation with an evolution time of 500. Inset: the size-dependence behavior of typical peaks with logarithm scale of vertical axis. }
\end{figure}

\begin{figure*}
\centering
\includegraphics[width=1\textwidth]{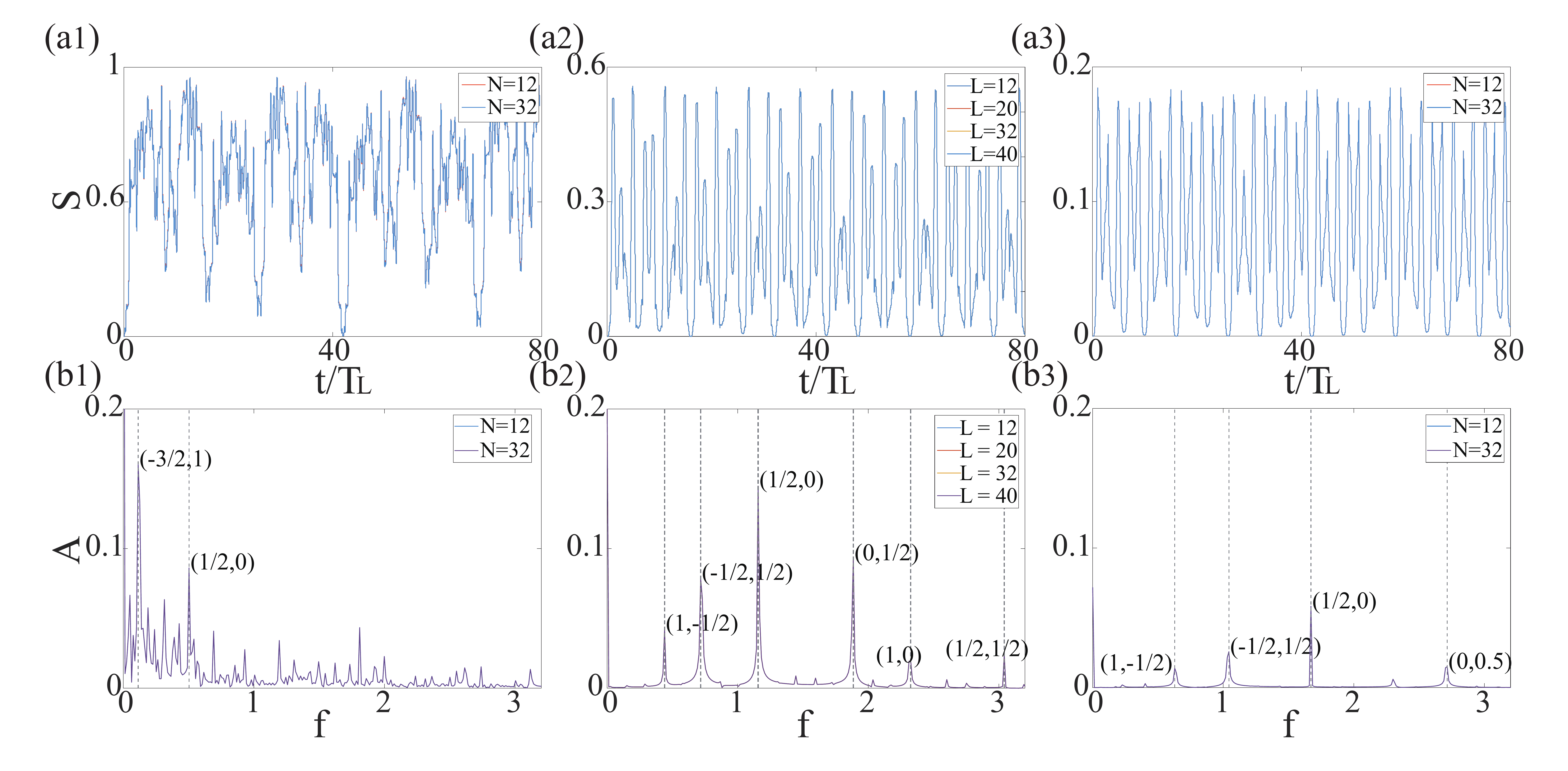}
\caption{\label{Fig4} The results of the EE at different driving frequencies. (a) The time-dependent behavior. (b) its associated Fourier spectrum. We performed Fourier transformation with an evolution time of 500. Here, we fix $\theta = \pi$. For (a1, b1) $f_{L} = 1.00$, (a2, b2) $f_{L} = 2.32$. (a3, b3) $f_{L} = 3.34$. Note $T_L$ also varies with the driving frequency.
In addition, we also plot the result of various system size with different colors.} 
\end{figure*}

\section{Fidelity}
We proceed to consider another observable to verify the presence of DTQC, which corresponds to the overlap with the initial $Z_2$ state. It thus directly expresses the recovery phenomenon to the initial state, and written as
\begin{equation}
    F = |\langle Z_2| \psi (t)\rangle|.
\end{equation}
Here we denote the observable as the fidelity in the time-evolution. As shown in Fig. \ref{Fig3}(a), the fidelity of the system also exhibits DTQC behavior within a certain phase region. For its size-dependence behavior, we find that as system size increases, the oscillation pattern has changed. Noticeably, the locations of Fourier peaks for different system sizes do not change. Meanwhile, we observe the typical peaks now contain (0, 0.5) and (0.5, 0) components, which contains the individual subharmonic response of each external drive, i.e. $f_L/2$ and $f_R/2$. Different from the result from antiferromagnetic observable $m$, the amplitude of Fourier peaks decrease as size increases. In the subfigure, we select two peaks (1/2, 1/2) and (-1/2, 1/2) to fit for its size-dependence amplitude, and find the exponentially decay behavior. This is interpreted as exponential increasement of the Hilbert space with respect to the system size $L$. Accordingly, the number of unconstrained states also get increased, and the overlap of the specific one (here we use $Z_2$ state) naturally decreases as the exponential function of the $L$. In the Appendix C, we also investigate the phase diagram of the fidelity in detail. We discover that such observable also manifests the behavior of DTQC and possesses certain robustness, with its associated phase diagram similar to Fig. \ref{Fig2}. Therefore, the observable of the fidelity also supports the emergence of DTQC behavior.

\section{Entanglement entropy}
Even though we have discussed the observed oscillatory behavior and its associated Fourier spectrum in the previous section, a general observable is required for further verifying that the DTQC is formed by the joint effect of two DTC phases. Here, we choose the bipartite entanglement entropy (EE) between the left and right region as the essential observable. Accordingly the EE is defined as $S = -\text{Tr}( \rho_{R} \log \rho_{R} ) $, with the reduced density matrix  $\rho_{R}$. We continuous to display the time-dependent evolution and its Fourier spectrum at different driving frequencies in Fig. \ref{Fig4}. The EE also demonstrates oscillatory phenomena, and we could also extract the distinguished peaks from its Fourier spectrum.  In Fig. \ref{Fig4}(a2), on the one hand, we find that the EE remains unchanged with the system size.
On the other hand, the bipartite EE has a relatively low value,  and maintains totally unthermalized due to Floquet modulation. Moreover, considering that the coupling between two bipartite comes merely from the Rydberg blockade effect of the nearest sites, we thus conclude that the entanglement between the left and right part is only confined to the boundary. Nonetheless, the non-zero value of EE still demonstrates that there exists the coupling between two DTCs by constructing the system Hamiltonian of Eq.\ref{PXPHamiltonian}. And we suppose that the DTQC appears by mixing the incommensurate response of two interacting DTCs.

Finally, we analysis the behavior of the phase at different driving frequencies $f$ in Fig. \ref{Fig4}(a). Deep into the dependence of driving frequency, we observe that the average value of EE and the oscillatory period are very sensitive to the driving frequency. When the driving frequency is relatively small, the system still exhibits stable oscillations, Besides, there is a considerable degree of entanglement, with the average value being around 0.8. 
At intermediate range, there are significant similarities of the Fourier spectrum between EE and the fidelity in Fig. \ref{Fig3}. We observe several clear peaks from the subharmonic response of each drive in Fig. \ref{Fig4}(b2), combined with its sum frequency and difference frequency. Such conditions could be identified as the proper parameter range for generating DTQC phase through EE criteria. 
When the driving frequency continues to increase, the entanglement becomes relatively small, and the amplitude of typical Fourier components also decrease. It shows that, under the fast oscillation of the atoms population, the Rydberg blockade effect becomes less obvious, and the two parts of the whole system gradually decouple. These findings are in accordance with the results in Fig. \ref{Fig2}(d). 

\section{Conclusion}
We report a proposal for realizing the DTQC phase in the Rydberg atomic chain. By coupling the two DTC phases via the Rydberg blockade effect, we obtain the stable mixture of the subharmonic response from both two phases with maximum incommensurability. In particular, we numerically observe distinguishable peaks at the Fourier spectrum, which satisfy the criteria for the DTQC phase. We also investigate the phase diagram and its features as a function of the tunable parameters to verify its robustness. To further describe the correlation between two original DTC phases, the bipartite entanglement entropy is employed, and a finite entanglement is identified close to the boundary. Such a phenomenon reveals that under a certain condition, the separated two subsystems could turn into a unity and exhibit the quasi-crystal behavior in the time domain. Our proposal thus provides the opportunity to construct novel phases on the popular Rydberg system. And the results are able to expand the understanding of the field of the quantum simulation, and the non-equilibrium phase of matter.

\section*{Acknowledgment}
We thank the helpful discussion with Heng Shen, Zhuangzhuang Tian. This work is supported by National Key R\&D Program of China under Grant No. 2020YFA0309400, NNSFC under Grant No. 12222409 and 12174081, 11974228, and the Key Research and Development Program of Shanxi Province (Grants No. 202101150101025). W. J. acknowledges National Natural Science Foundation of China (Project No. 12404275), and the fundamental research program of Shanxi province (Project No. 202403021212015). 

\section*{APPENDIX A: The conversion between DTC and DTQC}
\label{app:A}
\begin{figure}
\centering
\includegraphics[width=0.45\textwidth]{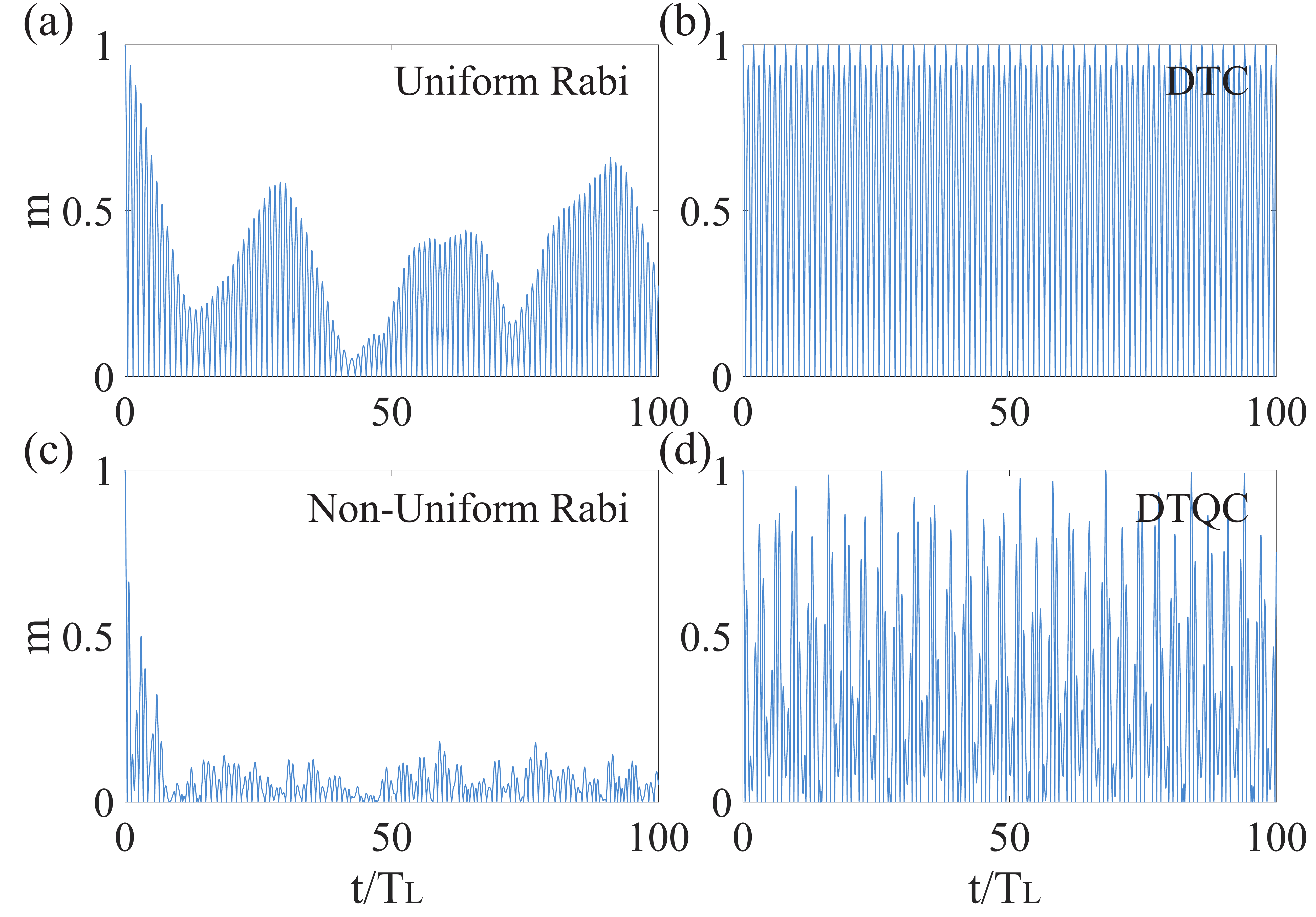}
\caption{\label{Fig5} The temporal evolution of the observable $m$ for constructing DTC and DTQC. The left two panels are the quenched behavior without the modulations, where (a) is the uniform Rabi case, (b) the non-uniform case.  The right two panels are applied with the  modulations, where (c) shows a DTC, and (d) shows a DTQC. (a) $\Omega_{R}/\Omega_{L}=1$, $\theta=0$. (b) $\Omega_{R}/\Omega_{L}=1$, $T_{L}=T_{R}=4.74$, $\theta=\pi$. (c) $\Omega_{R}/\Omega_{L}=(\sqrt{5}+1)/2$, $\theta=0$. (d) $\Omega_{R}/\Omega_{L}=(\sqrt{5}+1)/2$, $T_{L}=4.74$, $T_{R}=2T_{L}/(\sqrt{5}+1)$, $\theta=\pi$. The system size is chose as $N = 10$.}
\end{figure}

We conduct numerical simulations to study the temporal evolution of the observable $m$ under different parameters in Fig. \ref{Fig5}. When the Rabi frequency $\Omega$ of the atomic chain is uniform,  named $\Omega_L = \Omega_R$, the time evolution of the observable $m$ after quench will exhibit the scar phenomenon. Applying the global Floquet modulation enables $m$ to show stable subharmonic response. When the atomic chain is divided into two subsystems, in the absence of driving, $m$ will rapidly decay and the system will fleetly thermalize. With the addition of the double Floquet drive, $m$ shows a stable but quasi-periodic recovery.

\begin{figure}
\centering
\includegraphics[width=0.45\textwidth]{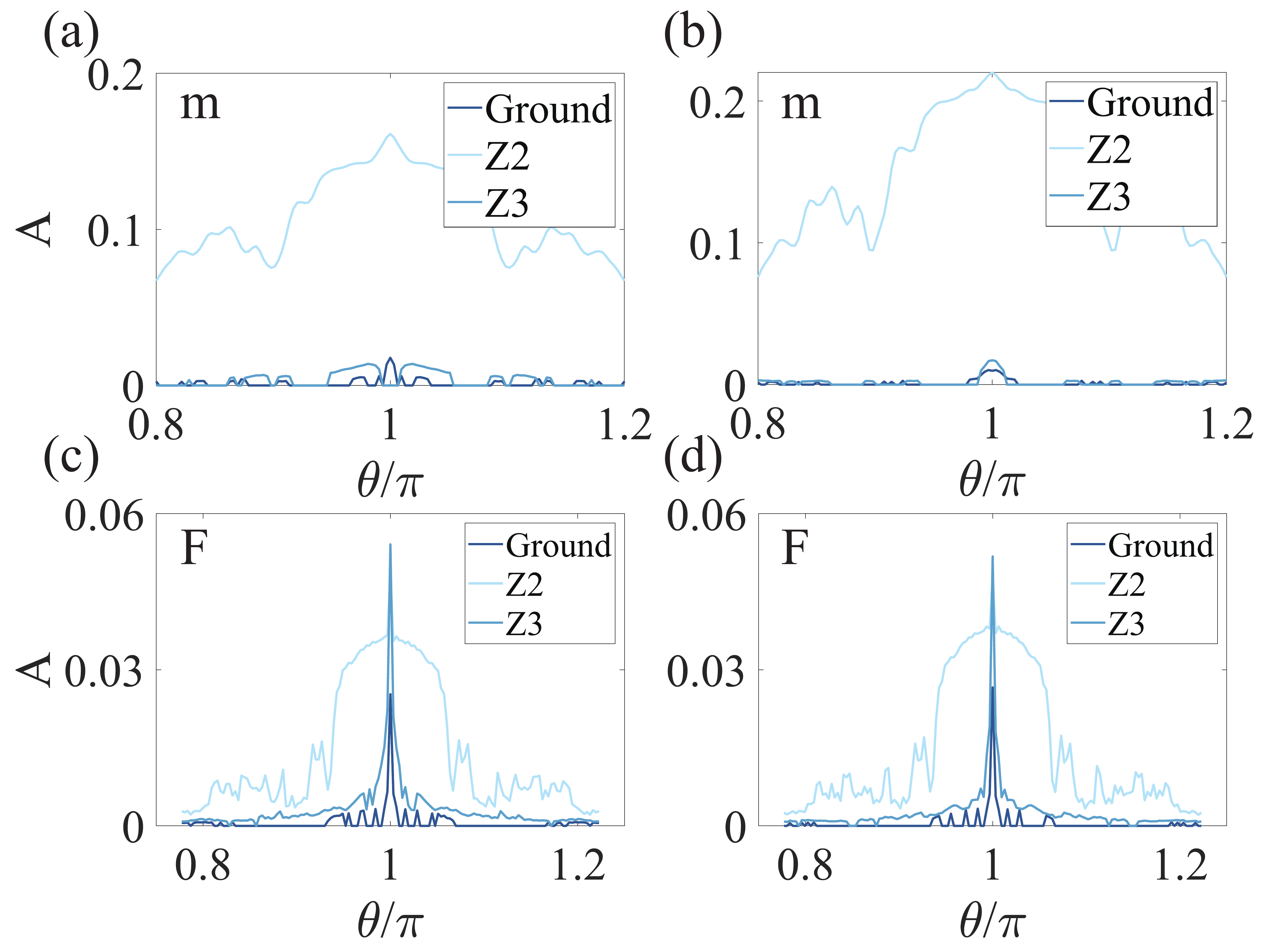}
\caption{\label{Fig6} Under different initial states, the peaks (-1/2, 1/2) and (1/2, 1/2) of the observable vary with the modulation intensity $\theta$. We choose $T_{L} = 4.74 $, the evolution time $t = 500$, and the system size $N = 12$. (a, b) The typical peaks at (1/2, -1/2) and (1/2, 1/2) for the observable $m$. (c, d) The typical peaks at (1/2, -1/2) and (1/2, 1/2) for the observable $F$.  }
\end{figure}

\section*{APPENDIX B: Initial state dependence}
\begin{figure}
\centering
\includegraphics[width=0.45\textwidth]{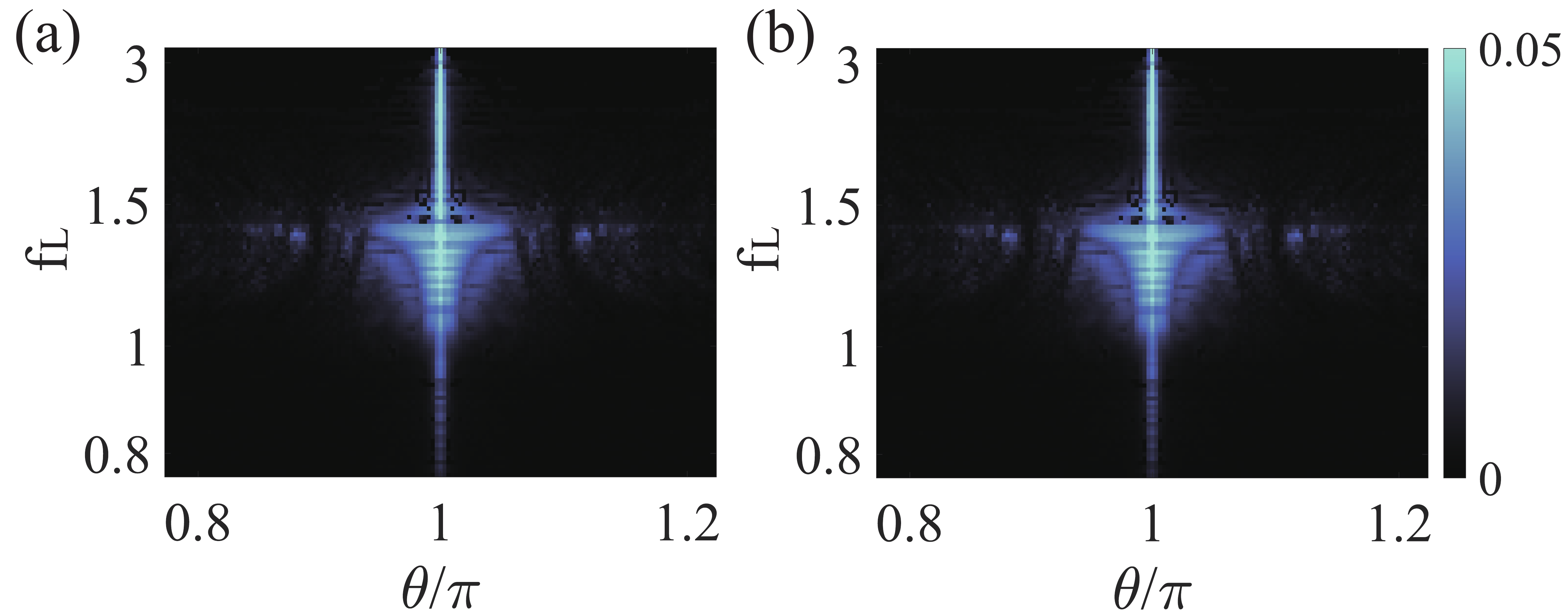}
\caption{\label{Fig7} Fidelity  $\langle Z_2| \psi (t)\rangle$ phase diagrams, where the criteria of the frequency components for (a) is (-1/2, 1/2) and (b) is (1/2, 1/2). We choose evolution time $t = 1000$, and system size $N = 10$.  }
\end{figure}

In this section, we also investigate the dependence of the observables $m$ and $F$ on the $Z_2$ state, the $Z_3$ initial state, and the state in which all atoms are in their ground state, which are three different initial states in Fig. \ref{Fig7}. Since only the $Z_2$ state has the $Z_2$ symmetry, it has a relatively stable trajectory in Hilbert space, which is known as the scar phenomenon. Through Floquet modulation, the trajectory becomes even more stable, and thus it shows greater robustness in both kinds of observables. For the other two initial states, when $\theta=\pi$, the observable $F$ of the three initial states can exhibit better oscillations. This is because the modulation effect can perfectly return to the initial state after two times for a single subsystem. However, when $\theta$ deviates slightly from $\pi$, both of the other two initial states will quickly become thermalized. In general, only when the initial state is the $Z_2$ state can robust oscillatory behavior occur.

\section*{APPENDIX C: Fidelity Phase Diagrams}
In Fig. \ref{Fig6}, We draw phase diagram of fidelity $F$ in Fig. \ref{Fig6}, which is similar to the observable $m$. However, the difference is that the range of DTQC phase for fidelity $F$ is narrower and the amplitude of the peaks is smaller. It is noted that when the modulation frequency $\theta=\pi$, there is a very good DTQC response for almost all modulation frequencies. This is because the modulation operator anticommutes with the Hamiltonian for subsystems, and subsystems perfectly evolve to the original state after two cycles. Compared with the observable $m$, fidelity can better reflect this phenomenon. See Ref. \cite{DTC} for its deep exploration. 

\bibliography{main}

\end{document}